\documentclass[12pt]{article}
\usepackage{amssymb}
\usepackage{graphicx}
\usepackage{amsmath}

\def\vsigma{\vec\sigma}

\def\vtau{\vec\tau}
\def\vnu{\vec\nu}

\def\<{\langle}
\def\>{\rangle}

\def\det{{\rm det}}

\def\dv{{\bf d}}
\def\jv{{\bf j}}
\def\rv{{\bf r}}

\def\Bv{{\bf B}}
\def\Ev{{\bf E}}

\def\zu{\hat{\bf z}}
\def\lambdabar{\lambda\raise0.4ex\hbox{\kern-0.5em\hbox{--}}\ }

\def\Fv{{\bf F}}

\def\F{{\cal F}}
\def\E{{\cal E}}
\def\B{{\cal B}}

\def\derx{\partial_X}
\def\dery{\partial_Y}

\def\Re{{\rm Re}\,}

\def\be{\begin{equation}}
\def\ee{\end{equation}}

\begin{document}

\begin{center}

{\Large Electron acceleration in a helical waveguide}%
\footnote{Presented at the VIII$^{\rm th}$ International Symposium \emph{Radiation from Relativistic Electrons in Periodic Structures} (RREPS-09), Zvenigorod, Russia, Sept. 7-11, 2009.}
\end{center}

\vspace{5mm}
\centerline{X. Artru, C. Ray}

\vspace{5mm}
\begin{center}
Institut de Physique Nucl\'eaire de Lyon, Universit\'e de Lyon,\\ Universit\'e Lyon 1 and CNRS/IN2P3, F-69622 Villeurbanne, France
\end{center}

\centerline{x.artru@ipnl.in2p3.fr, c.ray@ipnl.in2p3.fr}
\vspace{5mm}

\begin{abstract}
Travelling wave in a helical wave guide is considered for electron acceleration. A first determination of the travelling wave modes using a partial wave expansion (PWE) and a point matching method is presented. It gives a rapid solution for moderate deformation of the guide relative to a straight cylinder. Strong deformations will give higher accelerating gradient but the PWE diverges. Two methods overcoming this difficulties are suggested. 
\end{abstract}

\section{Introduction}
An electromagnetic wave in a straight wave guide cannot accelerate particles because its phase velocity $v_{\rm ph}$ is larger than $c$. In conventional linear accelerators, $v_{\rm ph}$ is reduced by a chain of resonant cavities along the guide. A scheme not using a resonance phenomenon is the travelling wave tube (TWT). One type of TWT is a cylindrical tube with a conducting helical wire inside \cite{HELIX,Aron:1973ya}. Another type is a tube with a thick dielectric wall \cite{MICA}. 

In this paper we study another kind of helical device~: the helically bent tube. The main "photon path" is the helix passing trough the centres of the local cross sections of the tube. For a strong enough bending, the difference in length between this path and a straight path is such that the effective $v_{\rm ph}$ is smaller than $c$, although the local $v_{\rm ph}$ is larger than $c$. 

One may expect several advantages in using helically bent tube in place of accelerating cavities~: easy construction, easy cooling, easy maintainance of the wall surface using an helical arm, large frequency spacing of the modes, large group velocity which allow a fast filling of the guide by the wave. Alternatively, the wave can be generated by a driving beam. Like with the dielectric wall in \cite{Kanareykin}, several modes can cooperate to transfer the energy from one beam to the other. 

The helical wave guide is defined by its axis $Oz$, its period $\Lambda=2\pi/q$ and the \emph{base curve}, which is the section of the wall with the $z=0$ plane, of parametric equation $\rv_b(s)=\left(x_b(s),y_b(s)\right)$. The point $(x,y)=(0,0)$ is supposed to be inside the base. The wall surface is generated by helices $H_s$ of parametric equation
\be\label{param}
(x,y,z)=\left({\cal R}[qz]\rv_b(s),\,z\right)
\ee 
where the $2\times2$ matrix ${\cal R}[qz]$ describes the rotation of angle $qz$ about $0z$.
One method to make a helical waveguide is to draw helical grooves on the internal face of an initially cylindral tube. The case of rectangular grooves has been studied in \cite{Wei,MAFIA}. The corresponding wall base curves have indentations. Here we will restrict to base curves with smooth curvatures. 

This work is the continuation of a first study \cite{helixErice}, where the electromagnetic wave modes were calculated using a partial wave expansion (PWE) associated to a point matching (PM) method. Since that time, we realized that PWE is limited to weak enough helical bending and sufficiently "round" base curves~; if it is not the case, the PWE converges only inside a cylinder which does not contain the whole guide. 

In this paper we will recall the PWE-PM method, some of our old results, present some new ones and discuss the limitations of this method. Two alternative methods will be evoked.


\section{Propagating modes in the scalar field case}

We first present the simpler case of a scalar field $\Phi(t,x,y,z)$, which obeys the Klein-Gordon equation (in units $c=1$) 
\be\label{scalar} 
(-\partial^2_t+ \partial^2_x+ \partial^2_y+ \partial^2_z)\, \Phi(t,x,y,z)=0\,,
\ee
with the boundary condition $\Phi(t,x,y,z)=0$ on the wall.
These two equations are invariant under a time translation and under a \emph{helical displacement}
\be \label{transf-helic}
\{\hbox {translation by }\Delta z\} \times 
\{\hbox {rotation by } q\Delta z\} \,.
\ee 
In the transverse plane we introduce a \emph{"rotating frame"}, the coordinates $(X,Y)$ of which are related to the fixed frame  ones by $x+iy = e^{iqz}\,(X+iY)$. 
On can therefore look for a solution of the form
\be\label{reduction-sca}
\Phi(t,x,y,z)=e^{iPz-i\omega t}\, \Psi(X,Y) \,
\ee
which has definite reduced frequencies $\omega$ and definite eigenvalue $P$ of the operator 
\be\label{Phelic} 
{\cal P}_{\rm helic}=p_z+q{\cal J}_z=-i\partial/\partial_z-iq\partial/\partial_\varphi
\,.\ee

\subsection{The partial wave expansion (scalar field case)}
Let us temporarily assume that a mode can be expanded in free cylindrical partial waves which are eigenstates of ${\cal J}_z$ and $p_z$ with eigenvalues $l$ and $p$ respectively. A single partial wave writes
\be\label{basic-scalaire} 
\Phi_{\omega,p,l}(t,x,y,z) = 
 e^{-i\omega t}\, e^{ip z}\, e^{il\varphi}\,\psi_{l,k}(r) \,,
\ee
where $(r,\varphi,z)$ are the cylindrical coordinates and
\be\label{Bessel}
\psi_{l,k}(r) = J_{|l|}(kr)/k^{|l|} = I_{|l|}(\kappa r)/\kappa^{|l|} 
\,.\ee
$\,J_l$ and $I_l$ are the regular Bessel functions of first and second kind, $k=(\omega^2-p^2+i0)^{1/2}$ is the radial momentum and $\kappa\equiv-ik=(p^2-\omega^2-i0)^{1/2}$. The two expressions of (\ref{Bessel}) are equivalent. If one whishes to work with real quantities, one takes the first one for $|p|\le|\omega|$, the second one for $|p|>|\omega|$. The denominator $k^{|l|}$ is introduced to make $\psi_{l,k}(r)$ finite at $k\to0$. The partial wave expansion reads
\be\label{decompose-scalaire} 
\Phi(t,x,y,z)=e^{-i\omega t}\,\sum_{l=-\infty}^{+\infty}  a_l \, e^{ip(l)z+il\varphi}\, \psi_{l,k(l)}(r) \,,
\ee
with
\be
p(l)\ {\rm or}\ p_l=P-lq\,,\quad k(l)\ {\rm or}\ k_l=\left(\omega^2-p_l^2+i0\right)^{1/2}
\,,\ee
so that each partial wave is an eigenstate of ${\cal P}_{\rm helic}$ with eigenvalue $P$.

It is sufficient to write the boundary condition at $t=z=0$, thanks to the invariances under time translation and helical displacement. For every point $\rv_b(s)$, of polar coordinates $(r_b(s),\varphi_b(s))$, of the base curve one has the constraint
\be
\sum a_l\, \exp[{il\varphi_b(s)}]  \, \psi_{l,k}[r_b(s)] =0 
\,.\ee

\subsection{The point matching method in the scalar field case}
\begin{figure}\label{courbase}
\begin{center}
\includegraphics[scale=0.4]{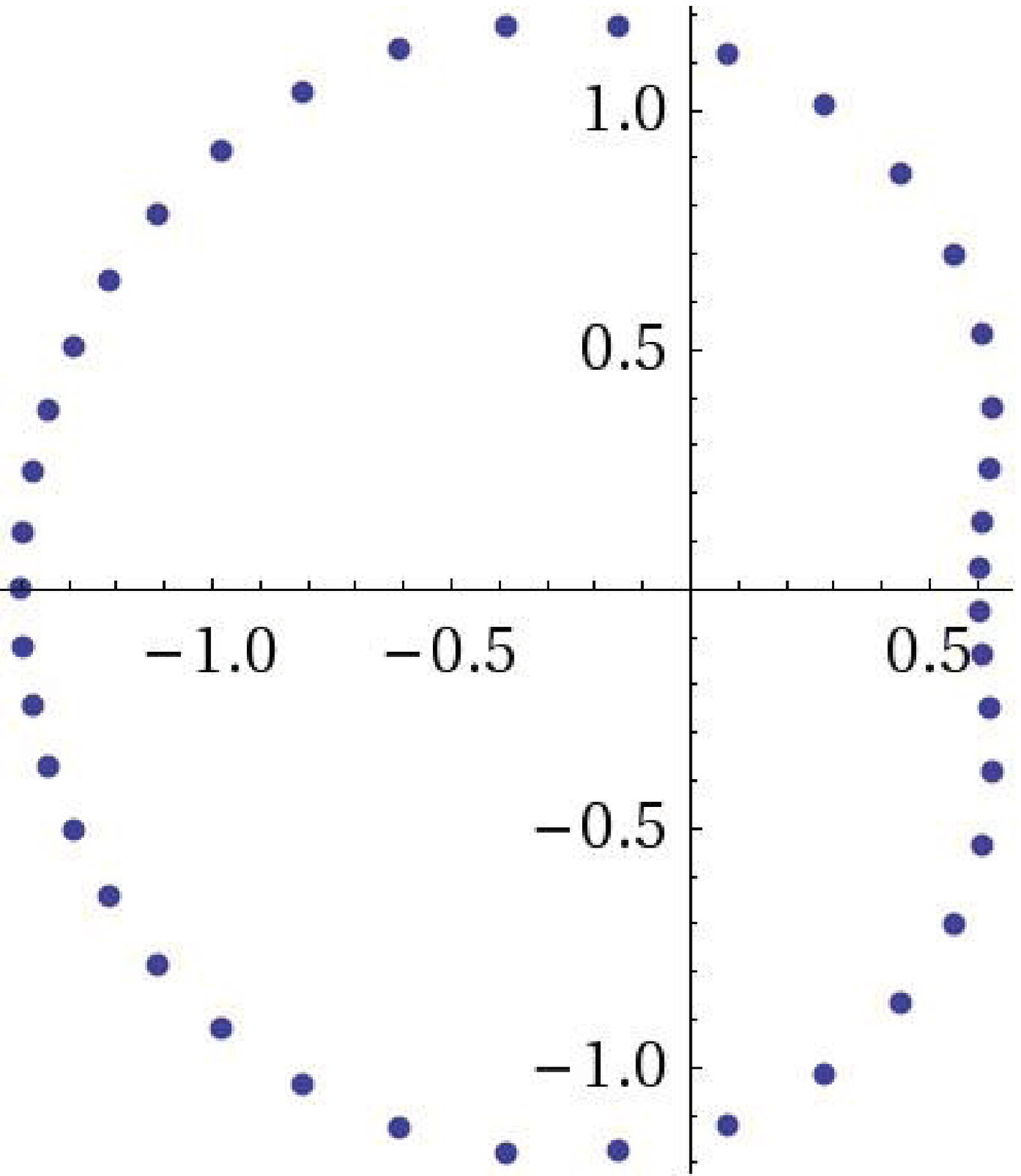} 
\end{center}
\caption{Matching points on the base curve.}
\end{figure}

One keeps only a finite set $\mathcal{L}$ of partial waves and selects an equal number of matching points $\rv_1,...\rv_N$ of the base curve (see Fig.\ref{courbase}). For a base without discrete rotational symmetry, $\mathcal{L}=[l_{\rm inf},l_{\rm sup}]$ with $l_{\rm sup}=l_{\rm inf}+N-1$. 
The choice of $\rv_n$'s and $\mathcal{L}$ should be guided by the following conditions~:
\begin{itemize}
	\item the $\rv_n$'s should give a precise enough definition of the base curve,
	\item $\mathcal{L}$ should contain $l$'s of both signs, however with more $l$'s of the same sign as $q$. 
	\item $l \,(\varphi_{n+1}-\varphi_n)$ for $l\in\mathcal{L}$ should not be too large compared to unity,
	\item $k_l \, (r_{n+1}-r_n)$ for $l\in\mathcal{L}$ should not be too large compared to unity. 
\end{itemize} 
The $a_l$ are approximatively determined by the system of linear equations 
\be\label{eq-lin-scal} 
\sum_{l\in\mathcal{L}} a_l\, e^{il\varphi_n} \, \psi_{l,k}(r_n) 
=0\,, \quad 1 \le n \le  N \,,
\ee
$\ (r_n,\varphi_n)$ being the polar coodinates of $\rv_n$. A nonzero solution exists if the determinant est zero:
\be\label{det-scal} 
\det\{C_{nl}\}=0 \,,\ {\rm where}\ \ C_{nl}= e^{il\varphi_n} \, \psi_{l,k}(r_n) \,.
\ee
This equation, which relates $\omega$ and $P$, is the dispersion relation in the PWE-PM approximation. One may impose $\omega$ and look for the roots of (\ref{det-scal}) as a function of $P$, or vice-versa. Alternatively, in view of accelerating highly relativistic electrons one may impose $\omega/P\equiv v_{\rm ph}=v_e\simeq1$ and look for the roots of (\ref{det-scal}) as a function of $\omega$.
Using a graphical software, the roots are easily seen as dips in the curve of $\ln|\det\{C_{nl}\}|$ versus the free variable. If $\left(\det\{C_{nl}\}\right)$ is $e^{i\alpha}$ times a real number, the roots are precisely located as jumps of the curve $\arg\left(ie^{-i\alpha}\,\det\{C_{nl}\}\right)$ from $\pm\pi/2$ to $\mp\pi/2$. 

Provided that the partial wave expansion converges, the quality of the approximation is improved by increasing $N$. The convergence is obtained first for the lowest modes (which have smallest $\omega$ at fixed $P$ or $v_{\rm ph}$, or largest $P$ at fixed $\omega$), then for higher and higher modes. A first criterium is the stability of the root when $N$ is increased by one or two units, when $\mathcal{L}$ is translated by one or two units or when one moves the matching points.
The ultimate criterium is the stability of the coefficients $a_l$, at least for $l$ not close to $l_{\rm inf}$ or $l_{\rm sup}$. 

\paragraph{Case of a symmetric base.} 

If the base is symmetrical about the $x$-axis, one can choose the $\{\rv_n\}$ to be symmetrical by pairs, except for those which are on the $x$-axis. Then $i^\nu\det\{C_{nl}\}$ is real, $\nu$ being the number of pairs. The $a_l$ can be taken all real, which gives 
\be \label{sym-0x} 
\Phi(0,x,-y,0)=\Phi^*(0,x,y,0)
\,.\ee

If the base is symmetrical about $Oz$, the modes are either even or odd under the parity symmetry $(x,y)\to(-x,-y)$, therefore $\mathcal{L}$ has to contain either even or odd $l$'s. 
More generally, if the base has a rotational symmetry of order $S$, the $l$'s should be equal modulo $S$. To avoid redundant equations in (\ref{det-scal}), the points $\rv_n$ should be all different modulo the symmetry. It suffices to choose them all on one sub-period of the base curve.  

\paragraph{Numerical exemple with the scalar field.} 
The simplest base curve is a circle or radius $a$ which is de-centered by  $\varepsilon a$ with respect to the $z$-axis. We took $\varepsilon=0.4$, $q=1$ and imposed the phase velocity $v_{\rm ph}=0.99999$. The lowest modes are at $\omega a=5.0$ and $7.0$. They are stable at the $0.1$ precision when $\mathcal{L}$ contains at least the set $[-1,+5]$. 
Fig.2 displays the cross section $\Psi(X,Y)$ of the field at $z=t=0$, for $\omega a=5.0$ mode. Note the symmetry (\ref{sym-0x}). The maximum of $|\Re\Psi|$ is not at the centre of the base, but farther from the helix axis. This is a manifestation of the centrifugal effect. 

\begin{figure}
\centering{%
\begin{tabular}{c c}
\includegraphics[width=.49\textwidth]{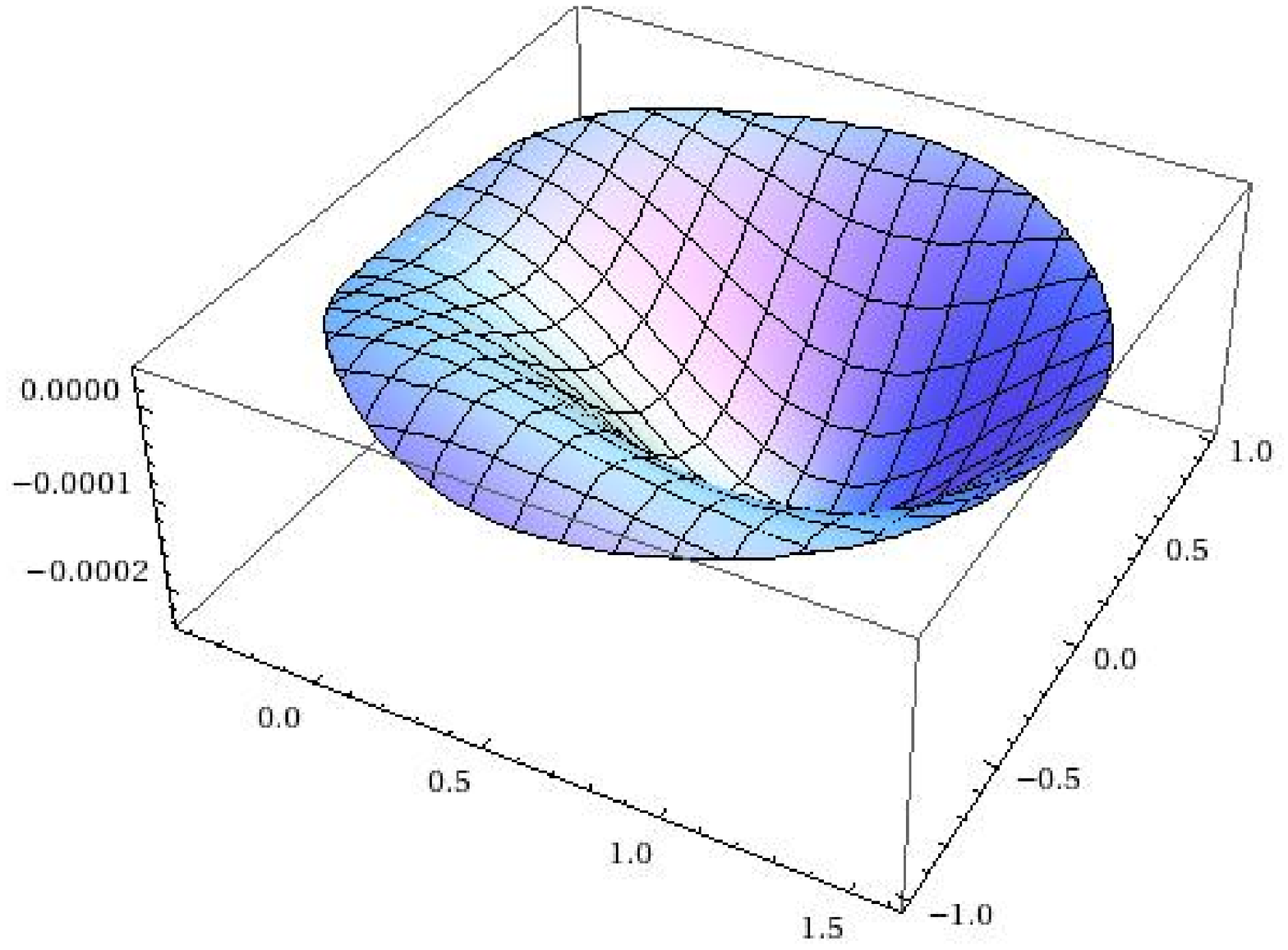} 
&
\includegraphics[width=.49\textwidth]{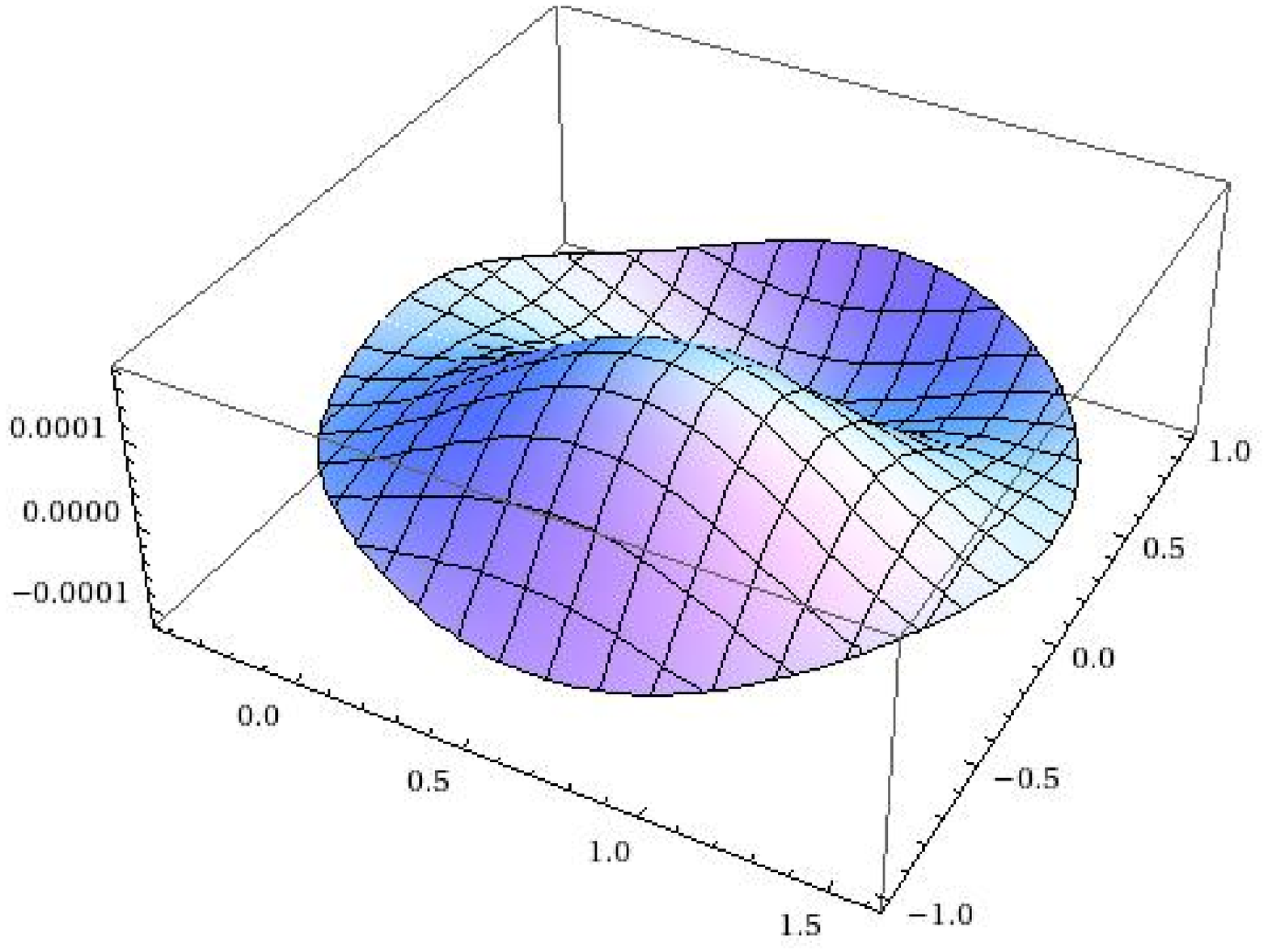} 
\end{tabular}
}
\caption{3-dimensional representation of the scalar field $\Psi(x,y)=\Phi(0,x,y,0)$ of the $\omega a=5.0$ mode, for a circular base of radius $a$, off-centred by $\varepsilon=0.4$ from the axis, and $q=1$. 
Left: real part; right: imaginary part.}
\end{figure}

\section{Propagating modes of the Maxwell field}

We gather the fields $\Ev$ and $\Bv$ in a 6-component vector $\Fv=\{ \Ev, \Bv \}$. 
In analogy with (\ref{reduction-sca}) we look for solutions of the form
\be\label{reduction} 
\Fv(t,x,y,z)=e^{iPz-i\omega t}\, {\cal R}[qz]\,\F(X,Y) 
\ee
with $\F(X,Y)=\{ \E(X,Y), \B(X,Y) \}$. ~${\cal R}[qz]$ is the rotation matrix of (\ref{param}) applied to the vectors $\E$ and $\B$.  The physical field is the real part of $\Fv$. 

\paragraph{Boundary conditions.} We introduce the vectors 
\be\label{tangent}
\vsigma(s)=\rv'_b(s)=\begin{pmatrix}x'_b(s)\cr y'_b(s)\cr 0 \end{pmatrix}, 
\quad \vtau(s)=\zu+q\,\zu\times\rv_b(s)=\begin{pmatrix}-qy_b(s)\cr qx_b(s)\cr 1\end{pmatrix},
\ee
which are tangent to the wall at $z=0$, and the normal vector $\vnu(s)=\vsigma(s)\times \vtau(s)$.
The boundary conditions are 
%
\be\label{boundary} 
\left\{\begin{array}{r}
\vec\E\cdot\vsigma=0  \\
\vec\E\cdot\vtau=0 \\
\vec\B\cdot\vnu=0
\end{array} \right.
\qquad\rightarrow\quad 
\left\{\begin{array}{r}
\E_r\,\sigma_r+\E_\varphi\,\sigma_\varphi =0 \\
qr\,\E_\varphi + \E_z =0 \\
\B_r \,\sigma_\varphi+(qr\B_z-\B_\varphi) \sigma_r =0
\end{array} \right.
\ee
The right set is written in cylindrical coordinates. These 3 conditions are not independent. Only two suffice. 

\paragraph{Partial wave expansion.}
$\Fv$ can be expanded in free partial waves like $\Phi$ in (\ref{decompose-scalaire}). However, since the photon has two polarisation states, there are two waves for each $l$, which now denotes the total (spin + orbital) angular momentum. We take  the states of right and left \emph{chirality}, or \emph{photon helicity} $\chi=\pm1$. The analogue of $\psi_{l,k}(r)$ in (\ref{basic-scalaire},\ref{decompose-scalaire}) is $\F_{\omega,p,l,\chi}(r)=
\left\{\vec\E_{\omega,p,l,\chi}(r),\vec\B_{\omega,p,l,\chi}(r)\right\}$. For $l=0$, the cylindrical components are 
\be\label{chiral-l=0}
\vec\E_{\omega,p,0,\chi}(r)=  \begin{pmatrix}
\E_r\cr \E_\varphi\cr \E_z\cr 
 \end{pmatrix}_{\omega,p,0,\chi} = \begin{pmatrix}
- ip\,\chi\,\psi_1(r)\cr 
\omega\,\psi_1(r)\cr
\chi\,\psi_0(r)\cr 
 \end{pmatrix}
 \,,\qquad \vec\B_{\omega,p,0,\chi}(r)=-i\chi\,\vec\E_{\omega,p,0,\chi}(r)
\,.\ee
%
For $l\ne0$, 
\be\label{chirale}
\vec\E_{\omega,p,l,\chi} = \begin{pmatrix}
-i \left[ \psi_{\lambda-1} + (p-\chi'\omega)^2 \, \psi_{\lambda+1}\right] \cr
s_l\,\left[\psi_{\lambda-1} - (p-\chi'\omega)^2 \, \psi_{\lambda+1}\right] \cr
2(p-\chi'\omega) \, \psi_{\lambda} \cr
 \end{pmatrix}
 \,,\qquad \vec\B_{\omega,p,l,\chi}=-i\chi\,\vec\E_{\omega,p,l,\chi}
\,.\ee
%
Here $s_l={\rm sign(l)}$, $\ \chi'=s_l\chi$, $\ \lambda=|l|$ and $\psi_{\lambda} = J_{\lambda}(kr)/k^{\lambda}$. 
These states have finite and  linearly independent limits when $p^2\to\omega^2$.
The analogue of (\ref{decompose-scalaire}) is 
\begin{eqnarray}
\label{decompose-vectoriel} 
\Fv(t,x,y,z)=
e^{-i\omega t}\,\sum_{l}  e^{il\varphi}\, e^{ip(l) z} \left[
a_l \, \F_{\omega,p(l),l,\chi=+s_l}(r) + b_l \, \F_{\omega,p(l),l,\chi=-s_l}(r) \right]
\,.
\end{eqnarray}
We made the choice to associate $a_l$ and $b_l$ with positive and negative $\chi'$ instead of $\chi$. For $l=0$ we set $s_l=+1$ in (\ref{decompose-vectoriel}), but we can alternatively take the (TM, TE) basis as in \cite{helixErice}.

\paragraph{The point matching method in the Maxwell field case.}

Generalising the PWE-PM method of Section 2.2 we keep only $N$ values of $l$ and write the first two conditions of (\ref{boundary}) for $N$ points $\rv_n$ of the base curve.  It gives $2N$ linear equations linking $N$ coefficients $a_l$ and $N$ coefficients $b_l$. The $2N\times2N$ matrix of the  system is given in \cite{helixErice}. The dispersion relation is the vanishing of its determinant. 

\paragraph{Numerical results for the Maxwell field.}

Taking the same de-centered circular base curve as in the scalar field case $\varepsilon=0.4$, but $q=1$ and imposing $v_{\rm ph}=0.999$, we obtain the lowest modes at $\omega a=2.28$, $3.6$, $4.8$ and $5.0$. They remain the same for various $\mathcal{L}$ sets like $[-6,+8]$, $[-5,+9]$ and $[-3,+11]$. 
The TM/TE character, measured by $|E_z(0)/B_z(0)|$, are 0.3, 0.25 and 0.4. and 5.4 respectively. The $\omega a=5.0$ mode seems therefore more appropriate for electron acceleration. It has a \emph{figure of merit} 
\be
f.o.m. = |E_z(0)|/[\max(|\Ev|) \text{ on the wall}] \simeq1/8\, 
\ee 
and a group velocity $v_{\rm g}=d\omega/dP=0.68$.
Plots of the fields of this mode can be seen in \cite{helixErice}. Let us mention that $|\B_r|$ and $|\B_\varphi|$ look very similar to $|\E_\varphi|$ and $|\E_r|$ respectively, which is reminiscent of a plane wave. 

Increasing the guide periods $\Lambda=2\pi/q$ while imposing fixed $v_{\rm ph}<1$ makes the $\omega$ of each mode increase until the mode disappear at $\omega=\infty$. For $q=0.7$, the two lowest modes are at $\omega a= 2.7$ and $4.2$ with $|E_z(0)/B_z(0)|=0.25$ and $0.26$ respectively. The lowest mode has $f.o.m\simeq1/9$. 

With a stronger helical bending of the waveguide (larger $\varepsilon$ and $q$), the mode frequencies decrease, but the convergence of the PWE-PM method becomes questionable. For $\varepsilon=0.6$, $q=1.5$ and $v_{\rm ph}= 0.999$, three first modes are seen at $\omega a=2.2$, $3.3$ and $4.0$ with $|E_z(0)/B_z(0)|$ = 0.35, 0.3 and 0.8 respectively. About twenty matching points were sufficient to obtain stable roots. However the field amplitudes were not stable. 

We also considered a non-circular base curve of the Pascal's lima\c con type (figure 1), parametrized by
\be
r_b=a\,(1-\varepsilon\cos s)\,,\quad \varphi_b=s-\varepsilon'\sin s\,,\quad s\in[0,2\pi]
\,.\ee
The inward bump is expected to give a stronger $E_z(0)$. Indeed when a large part of the base curve is close to the origin, the boundary condition inhibits $E_z(0)$. This is avoided with the bump. 
For $\varepsilon=\varepsilon'=0.3$, $q=0.7$ and $v_{\rm ph}= 0.999$, two modes are clearly seen at $\omega a=2.4$ and $6.7$. A state at $\omega a=3.5$ was also seen for odd $l_{\rm sup}$, but it appeared only as a smooth dip of $\ln|\det\{C_{nl}\}|$ for even $l_{\rm sup}$. 

Finally we considered a base curve with two inside bumps and a rotational symmetry of order 2, paramerized by
\be
r_b=a\,(1-\varepsilon\cos s)\,,\quad \varphi_b=(s-\varepsilon'\sin s)/2\,,\quad s\in[0,4\pi]
\,.\ee
The modes are either even or odd under this symmetry. Odd modes are not interesting for particle acceleration since $E_z(0)=0$. Even modes have the property $E_x(0)=E_y(0)=0$, which guaranties that electrons on the $z$-axis are not deflected and emit no synchrotron radiation. 
For $\varepsilon=\varepsilon'=0.1$, $q=0.7$ and $v_{\rm ph}= 0.999$, even modes are seen at $\omega a=4.0$ and $6.3$, with the different $l$ truncations $[l_{\rm inf},l_{\rm sup}] = [-4,8]$, $[-8,8]$ and $[-10,10]$. With more pronounced bumps ($\varepsilon=\varepsilon'=0.2$ or 0.3), no stable root was found. 

\section{Limits of the PWE-PM method}

We have encountered difficulties when the deformation of the wave guide relative to a straight cylindrical tube becomes too strong. This may be attributed not to the point matching method but to a lack of convergence of the partial wave expansion at $r$ larger than a critical radius $r_c$. 

Let us indeed consider the expansion (\ref{decompose-scalaire}), in the scalar field case, and suppose that it converges at some point of polar coordinates $(r_0,\varphi_0)$. Then it converges at all points $(r,\varphi)$ with $r<r_0$. It implies that the domain of convergence is bounded by a cylinder. 

\paragraph{Demonstration :}

The convergence of (\ref{decompose-scalaire}) at $(r_0,\varphi_0)$ implies that $|a_l\, \psi_{l,k(l)}(r_0)|$ is bounded by some number $M$. Using (\ref{Bessel}), we have 
\be
|a_l\, \psi_{l,k(l)}(r)|\le M\,
\left|{I_{|l|}(\kappa_l r)\over I_{|l|}(\kappa_l r_0)}\right|
\ee
for any $r$. At large $|l|$, we have 
$\kappa_l 
\simeq |l|q-s_lP$ with $s_l={\rm sign}(l)$ and the asymptotic form of the modified Bessel functions gives $I_{|l|}(\kappa_l r)/I_{|l|}(\kappa_l r_0)\simeq \exp[(|l|q-s_lP)\,(r-r_0)]$. Then 
\be
|a_l\, \psi_{l,k(l)}(r)|\le M\,\exp[s_lP(r_0-r)]\,e^{|l|q(r-r_0)} \quad\text{at large }l
\,.\ee
If $r<r_0$, the expansion (\ref{decompose-scalaire}) is bound in modulus by a convergent geometric series in $|l|$, therefore converges at $l\to\pm\infty$. 

\medskip
If the base curve has a singularity, \emph{e.g.,} an anglar point of polar coordinate $(r_1,\varphi_1)$, the field is  non-analytic at this point, therefore $r_c\le r_1$. On the other hand, for a strong enough deformation relative to the straight cylinder, part of the wave guide may be outside the cylinder of convergence, even if the base curve is analytic. A similar situation occurs when one uses the Rayleigh expansion for the wave diffracted by a grating~: this expansion does not converge down to the bottom of the grooves if these ones are too deep \cite{Petit-C,Hill-C,Agassi-G}. 

\section{Conclusion}


We have seen that an increase in the strength of the helical bending (increase of $\varepsilon$ and $q$ in the case of a de-centered circular basis) leads to a decrease of the dimensionless frequencies $\omega a$ of the low modes at imposed phase velocity $v_{\rm ph}<c$ ($a$ characterizes the transverse size of the guide). One can also expect that it leads to a larger accelerating field $E_z(0)$, at imposed power of the wave or imposed maximum field on the wall. We have indeed, in order of magnitude
\be
E_z/E_T \sim p_T/p_z\,,\quad p_T\sim 1/a\,,\quad p_z\sim\omega  \quad\to\quad E_z/E_T \sim1/(\omega a)
\,.\ee
Inward bumps of the base curve also favor large $E_z(0)$. 
On the other hand, the PWE-PM method is limited to weak helical bending and sufficiently "round" base curves. Here we propose two numerical methods to find the propagating modes for larger bending and base curves with more pronounced structures. 

\paragraph{The finite element method.}
This is the standard way to solve numerically the Maxwell equations. In the rotating frame these equations take the 2-dimensional form at fixed $\omega$ and $P$~:
\begin{eqnarray}\label{equadif}  
i\omega\B_X&=&+\dery\E_z-iP\E_Y-q\E_X+q(X\dery-Y\derx)\E_Y\,,\cr
i\omega\B_Y&=&-\derx\E_z+iP\E_X-q\E_Y-q(X\dery-Y\derx)\E_X\,,\cr
i\omega\B_Z&=&+\derx\E_Y-\dery\E_X\,,\cr
-iP\B_Z&=&\derx\B_X+\dery\B_Y+q(Y\derx-X\dery)\B_z
\,,
\end{eqnarray}
and analogous equations obtained with the duality transformation $\B\to\E$, $\E\to-\B$. 

\paragraph{Method of virtual currents.}
Instead of expanding the field in cylindrical partial waves, one does it with eigenfunctions of ${\cal P}_{\rm helic}$ (equation \ref{Phelic})which are regular inside the waveguide but singular at some outside places. A candidate is the retarded or advanced field produced by the charge-current distribution
\be
j^\mu(t,\rv,z)=
\begin{pmatrix} \rho \\ \jv_T \\ j_z \end{pmatrix}
=e^{iPz-i\omega t}\,\delta^{(2)}\big(\rv-{\cal R}[qz]\dv\big)\,
\begin{pmatrix} 1
\\ qv_{\rm ph}\,\zu\wedge\rv
\\ v_{\rm ph}
\end{pmatrix}
\,,\ee
with $v_{\rm ph}=\omega/P$. ~${\cal R}[qz]$ is the rotation matrix of (\ref{param}).
The $x$ and $y$ components of any vector are gathered in a bi-vector. The point $(x,y,z)=\left({\cal R}[qz]\dv, z\right)$ draws an helix of path $\Lambda=2\pi/q$ intersecting the $z=0$ plane at point $(\dv,0)$. ~$\rho(t,x,y,z)$ is a periodic distribution of charges running along this helix with the longitudinal velocity $v_z= v_{\rm ph}$ and  $\jv$ is the associated current. 
$j^\mu$ is periodic in time with frequency $\omega$ and is an eigenfunction of ${\cal P}_{\rm helic}$ with eigenvalue $P$, therefore the retarded or advanced field has the same property. 
Let us denote, for instance, by $\Fv^{(\dv)}=\{\Ev^{(\dv)},\Bv^{(\dv)}\}$ the retarded field. Different $\dv$'s give different independent solutions. To take into account the fact that there are two polarisations, we also consider the dual field $\tilde{\Fv}^{(\dv)}=\{\Bv^{(\dv)},-\Ev^{(\dv)}\}$, which would be produced by a distribution of magnetic charge and current. In the $z=0$ plane these fields have a singularity in $1/|\rv-\dv|$ at $\rv=\dv$ ; they are Green's functions of (\ref{equadif}).
Choosing $N$ singular points $\{\dv_1, \dv_2, \cdots \dv_N, \}$ external to the base curve, we may approximate the field of a mode by 
\be 
\Fv=\sum_{n=1}^N a_n \, \Fv^{(\dv_n)} + b_n \, \tilde{\Fv}^{(\dv_n)}
\,,
\ee
The coefficients $a_n$ and $b_n$ should be choosen to satisfy the boundary conditions on $N$ matching points $\rv_1,...\rv_N$ of the base curve. 

Near a sharp inward bump of the base curve the field is strong and, by analytic continuation, one probably meets a singularity not far behind the curve,  analogous to a point-like electrical image in electrostatics. This singularity may be simulated by placing one or a few $\dv_n$'s on it or near to it. It amounts to replace the continuous surface charge and currents on the bump by charges and currents on virtual helical wires. 

Whichever method will be used, it will be usefull to compare it with the PWE-MP method in the case of weak helical deformation, as a check.

\end{document}